# Angular Analysis of $\Lambda_b$ Decays into $\Lambda V(1^-)$

## Applications to Time-Odd Observables and CP Violation in $\Lambda_b$ Decays (I)

Z.J.Ajaltouni[1], E.Conte

*Laboratoire de Physique Corpusculaire de Clermont-Ferrand*
*IN2P3/CNRS Université Blaise Pascal*
*F-63177 Aubière Cedex FRANCE*

**Abstract**

A complete study of the angular distributions of the processes $\Lambda_b \to \Lambda V(1^-)$ with $\Lambda \to p\pi^-$ and $V(J/\Psi) \to \ell^+\ell^-$ or $V(\rho^0,\omega,\phi) \to \pi^+\pi^-$, $K^+K^-$ is performed. Emphasis is put on the initial $\Lambda_b$ polarization produced in the proton-proton collisions and, *without any dynamical assumption*, polarization density-matrices of the vector-mesons V are derived and help to construct T-odd observables which allow us to perform tests of both Time-Reversal and CP violation.

---

[1]email: ziad@clermont.in2p3.fr

# 1 Introduction

With the advent of B factories at the proton-proton colliders, huge statistics of beauty hadrons are expected to be produced which allow a thorough study of CP violation processes with the B mesons. On the other hand, some specific phenomena related to either b-quark physics or CP violation can be performed to put limits on the validity of the Standard Model. One of these processes concerns the validity of the Time Reversal (TR) symmetry and a promising method to look for TR violation is the *three body* $\Lambda_b$ decay [1], [2] as it was initiated with the hyperons long time ago [3].

⋆ T-odd operator is derived from Time Reversal but it keeps the initial and final states *unchanged*. It is well known that the time reversed state of a decay like $\Lambda \to p\pi^-$ or $\beta$ nucleon decay cannot be realized in the physical world and thus, we must be contented with the following transformations which are the main ingredients of TR operator :

$$\vec{r} \to \vec{r} \,,\ \vec{p} \to -\vec{p} \,,\ \vec{\ell} \to -\vec{\ell} \,,\ \vec{s} \to -\vec{s}$$

where $\vec{\ell}$ and $\vec{s}$ are respectively the angular momentum and the spin of any particle with momentum $\vec{p}$. Consequently the helicity of the particle defined by $\lambda = \vec{s}.\vec{p}/p$ remains unchanged by TR.

⋆ In the past, it was pointed out by many authors [4] the importance to look for T-odd effects in the hyperon decays like $\Lambda, \Sigma,$ and $\Xi$, as a consequence of both *CPT theorem* and CP violation in weak $|\Delta S| = 1$ decays. As far as beauty hadrons $\Lambda_b, \Sigma_b, \Xi_b$ are concerned, because of their numerous decay channels and the strength of CP violation within the b-quark family, opportunities to find T-odd observables will increase and interesting tests of both SM predictions and models beyond the SM can be performed successfully.
T-odd observables can be constructed from the decay products like $\vec{v_1} \cdot (\vec{v_2} \times \vec{v_3})$ where $\vec{v_i}$ is either the spin or the momentum of particle $\imath$. These observables change sign under TR and a non-vanishing mean value of their distribution could be a sign of TR violation.

This paper is the first of a series of two papers (or more) devoted to an exhaustive study and the *simulations* of $\Lambda_b$ decays into $\Lambda\ell^+\ell^-$ and $\Lambda h^+h^-$. Final leptons $\ell = e, \mu$ or final hadrons $h = \pi, K$ can originate from intermediate resonances which quantum numbers are those of a vector meson $1^-$ like $J/\psi, \rho^0, \omega$ and $\phi$. The case of a non-resonant state decaying into $\ell^+\ell^-$ via electroweak penguin diagrams is also envisaged and will be discussed further.

• In the present paper, emphasis is put on the *angular distributions* of both the intermediate states and the final particles in some appropriate frames, the helicity frames. By stressing on the importance of the polarizations of the initial $\Lambda_b$ and the intermediate resonances, complete calculations based on the helicity formalism are performed and take into account the spin properties of the final decay products. No dynamical assumption is made in these calculations which only use Relativistic Kinematics and the standard principles of Quantum Mechanics.
• The evaluations of the hadronic matrix elements appearing in the analytic expressions are based on the Operator Product Expansions (OPE) techniques and they are left for the second paper which is under preparation [5].

## 2 Decay of $\Lambda_b$ into $\Lambda V$ ($V = 1^-$)

In the collisions $pp \to \Lambda_b + X$, the $\Lambda_b$ is produced with a *transverse polarization* in a similar way than the ordinary hyperons; its longitudinal polarization is suppressed because of parity conservation in Strong Interactions.

Let us define $\vec{N}_P$ the vector normal to the production plane by the relation :

$$\vec{N}_P = \frac{\vec{p_1} \times \vec{p_b}}{|\vec{p_1} \times \vec{p_b}|}$$

where $\vec{p_1}$ and $\vec{p_b}$ are respectively the vector-momentums of one incident proton beam and the $\Lambda_b$. The mean value of the $\Lambda_b$ spin along $\vec{N}_P$ is the $\Lambda_b$ transverse polarization and usually it is greater than 20% [6].

Let ($\Lambda_b xyz$) be the rest frame of the $\Lambda_b$ particle. The quantization-axis $\Lambda_b z$ is chosen parallel to $\vec{N}_P$; the other orthogonal axis $\Lambda_b x$ and $\Lambda_b y$ are chosen arbitrarily in the production plane; particularly $\Lambda_b x$ axis can be taken parallel to the momentum $\vec{p_1}$.
The spin projection of the $\Lambda_b$ along the transverse axis $\Lambda_b z$ is designated by $M_i$ which value could be $\pm 1/2$. The *polarization density matrix* (PDM) of the $\Lambda_b$ is designated by $(\rho^b)$. It is a $(2 \times 2)$ hermitean matrix and its elements $\rho_{ij}$ verify the standard relations :
⋆ $\rho_{++}$ and $\rho_{--}$ are real with $\rho_{++} + \rho_{--} = 1$.
⋆ $\rho_{+-} = \rho_{-+}^*$
The physical meaning of some matrix elements is straightforward :
$\rho_{++}$ and $\rho_{--}$ represent the *probability* that the $\Lambda_b$ is produced with $M_i = +1/2$ and $M_i = -1/2$ respectively. So, the initial $\Lambda_b$ polarization is given by :

$$< \vec{S}_{\Lambda_b} \cdot \vec{N}_P > = \mathcal{P}_{\Lambda_b} = \rho_{++} - \rho_{--} \quad (1)$$

In the decay $\Lambda_b(M_i) \to \Lambda(\lambda_1) V(\lambda_2)$, $\vec{p} = (p, \theta, \phi)$ is the vector-momentum of the hyperon $\Lambda$ in the $\Lambda_b$ frame (fig.1); while $\lambda_1$ and $\lambda_2$ are the respective helicities of $\Lambda$ and $V$ with the possible values $\lambda_1 = \pm 1/2$ and $\lambda_2 = -1, 0, +1$. On the other hand, the momentum projection along the ($\Delta$) axis (parallel to $\vec{p}$) is given by :
$M_f = \lambda_1 - \lambda_2 = \pm 1/2$.
The $M_f$ values *constrain* those of $\lambda_1$ and $\lambda_2$ since, among six combinations, only *four* are physical. They are summarized in table 1 :

| $\lambda_1$ | $\lambda_2$ | $M_f$ |
|---|---|---|
| 1/2 | 1 | -1/2 |
| 1/2 | 0 | 1/2 |
| -1/2 | -1 | 1/2 |
| -1/2 | 0 | -1/2 |

Table 1: Possible helicity values respectively for $\Lambda$ and $V$ hadrons

The analytic form of the decay amplitude is obtained by applying the Wigner-Eckart theorem to the S-matrix element in the framework of the Jacob-Wick helicity formalism [7] :

$$A_0(M_i) = <1/2, M_i|S|p, \theta, \phi; 1/2, 1; \lambda_1, \lambda_2> = M_{\Lambda_b}(\lambda_1, \lambda_2) D^{1/2\star}_{M_i M_f}(\phi, \theta, 0) \quad (2)$$

- The hadronic matrix element $M_{\Lambda_b}(\lambda_1, \lambda_2)$ which will be denoted afterwards by $\Lambda_b(\lambda_1, \lambda_2)$ for simplicity contains all the *decay dynamics*. It will be explicitely calculated in the next paper [5].
- The Wigner matrix element is expressed according to the Jackson convention :

$$D^j_{M_i M_f}(\phi, \theta, 0) = d^j_{M_i M_f}(\theta) \exp(-iM_i \phi) \quad (3)$$

## 3  Decay of the intermediate resonances

By performing appropriate rotations and Lorentz boosts, we can study the decay of each resonance in its own helicity frame such that the quantization axis is parallel to the resonance momentum in the $\Lambda_b$ frame :

$$\overrightarrow{O_1 z_1} \parallel \vec{p_\Lambda} \quad \text{and} \quad \overrightarrow{O_2 z_2} \parallel \vec{p_V} = -\vec{p_\Lambda}$$

So, the values of the resonance spin projections along these axis are their *respective helicities* in the $\Lambda_b$ frame; $m_1^0 = \lambda_1$, $m_2^0 = \lambda_2$.

Considering the following decays :

$$\Lambda(\lambda_1) \to P(\lambda_3) \pi^-(\lambda_4)$$

and

$$V(\lambda_2) \to \ell^-(\lambda_5)\, \ell^+(\lambda_6) \quad \text{or} \quad V(\lambda_2) \to h^-(\lambda_5)\, h^+(\lambda_6)$$

the respective helicities of the final particles are :
⋆ $\lambda_3 = \pm 1/2$, $\lambda_4 = 0$
⋆ $\lambda_5, \lambda_6 = \pm 1/2$ if leptons or $\lambda_5 = \lambda_6 = 0$ if $0^-$ mesons.

In the $\Lambda$ helicity frame, the projection of the total angular momentum along the proton momentum $\vec{p_P}$ is given by $m_1 = \lambda_3 - \lambda_4 = \pm 1/2$, while in the vector meson helicity frame this projection is equal to $m_2 = \lambda_5 - \lambda_6 = -1, 0, +1$ if leptons or $m_2 = 0$ if $\pi$ or $K$. Departing from these relations the decay amplitude of each resonance can be written again a la Wigner-Eckart, requiring that the kinematics of its decay products (emission angles and helicities) are fixed. So,

$$A_1(\lambda_1) = <\Lambda(\lambda_1)|S^{(1)}|p_1, \theta_1, \phi_1; \lambda_3, \lambda_4> = M_\Lambda(\lambda_3, \lambda_4) D^{1/2\star}_{\lambda_1 m_1}(\phi_1, \theta_1, 0) \quad (4)$$

$$A_2(\lambda_2) = <V(\lambda_2)|S^{(2)}|p_2, \theta_2, \phi_2; \lambda_5, \lambda_6> = M_V(\lambda_5, \lambda_6) D^{1\star}_{\lambda_2 m_2}(\phi_2, \theta_2, 0) \quad (5)$$

where the hadronic matrix elements $M_\Lambda(\lambda_3, \lambda_4)$ and $M_V(\lambda_5, \lambda_6)$ will be denoted respectively by $\Lambda(\lambda_3, \lambda_4)$ and $V(\lambda_5, \lambda_6)$.

## 4  Analytical form of the decay probability

The relations written above allow us to expresss a general decay amplitude for the process :

$$\Lambda_b(M_i) \to \Lambda(\lambda_1)\, V(\lambda_2) \Longrightarrow P\pi^-\, \ell^+\ell^-\, (h^+ h^-)$$

⋆ This decay amplitude must include all the possible intermediate states, thus a sum over the helicity states $(\lambda_1, \lambda_2)$ must be performed.

$$\mathcal{A}_I = \sum_{\lambda_1, \lambda_2} A_0(M_i) A_1(\lambda_1) A_2(\lambda_2) \tag{6}$$

Using the explicit relations (1)-(4), the decay amplitude gets the form :

$$\begin{aligned}\mathcal{A}_I(\lambda_3, \lambda_4, \lambda_5, \lambda_6) &= \sum_{\lambda_1, \lambda_2} \Lambda_b(\lambda_1, \lambda_2) d^{1/2}_{M_i M_f}(\theta) \exp(iM_i\phi) \\ &\times \Lambda(\lambda_3, \lambda_4) d^{1/2}_{\lambda_1 m_1}(\theta_1) \exp(i\lambda_1\phi_1) V(\lambda_5, \lambda_6) d^{1}_{\lambda_2 m_2}(\theta_2) \exp i(\lambda_2\phi_2)\end{aligned} \tag{7}$$

where $\theta_1$ and $\phi_1$ are respectively the polar and azimuthal angles of the proton momentum in the $\Lambda$ rest frame while $\theta_2$ and $\phi_2$ are those of $\ell^-(h^-)$ in the $V$ rest frame.

⋆⋆ In a second step, we must take the squared modulus of the amplitude above in order to compute the decay probability. Since the $\Lambda_b$ spin component $M_i$ is *unknown*, the polarization density-matrix $(\rho^b)$ must be used in the following form :

$$d\sigma \propto \sum_{M_i, M'_i} \rho_{M_i M'_i} \mathcal{A}_I \mathcal{A}_I^* \tag{8}$$

where $M_i, M'_i = \pm 1/2$, the other parameters defining the final state being fixed.

⋆⋆⋆ In a third step, we must take account of the fact that the helicities of the final particles *are not measured and summation over the helicity values* $\lambda_3, \lambda_4, \lambda_5$ *and* $\lambda_6$ *must be performed* :

$$d\sigma \propto \sum_{\lambda_3, \lambda_4 \lambda_5, \lambda_6} \sum_{M_i, M'_i} \rho_{M_i M'_i} \mathcal{A}_I \mathcal{A}_I^* \tag{9}$$

Taking the explicit form of the amplitude $\mathcal{A}_I$, arranging the different terms which appear in the relation above and noticing that $\lambda_4 = 0$, the decay probability takes the following form :

$$d\sigma \propto \tag{10}$$

$$\sum_{\lambda_1, \lambda_2, \lambda'_1, \lambda'_2} \sum_{M_i, M'_i} \rho_{M_i M'_i} \Lambda_b(\lambda_1, \lambda_2) \Lambda_b^*(\lambda'_1, \lambda'_2) d^{1/2}_{M_i M_f}(\theta) d^{1/2}_{M'_i M'_f}(\theta) \exp i(M_i - M'_i)\phi$$

$$\times \sum_{\lambda_3} |\Lambda(\lambda_3, 0)|^2 d^{1/2}_{\lambda_1 m_1}(\theta_1) d^{1/2}_{\lambda'_1 m_1}(\theta_1) \exp i(\lambda_1 - \lambda'_1)\phi_1$$

$$\times \sum_{\lambda_5, \lambda_6} |V(\lambda_5, \lambda_6)|^2 d^{1}_{\lambda_2 m_2}(\theta_2) d^{1}_{\lambda'_2 m_2}(\theta_2) \exp i(\lambda_2 - \lambda'_2)\phi_2$$

where $m_1 = \lambda_3 - \lambda_4 = \pm 1/2$ and $m_2 = \lambda_5 - \lambda_6$. Summing over the helicity variables of the final decay products and recalling that $M_f = \lambda_1 - \lambda_2$, $M'_f = \lambda'_1 - \lambda'_2$, a more compact form is obtained such that only the intermediate resonance helicities appear :

$$d\sigma \propto \sum_{\lambda_1,\lambda_2,\lambda'_1,\lambda'_2} D_{\lambda_1-\lambda_2,\lambda'_1-\lambda'_2}(\theta,\phi)\Lambda_b(\lambda_1,\lambda_2)\Lambda_b^*(\lambda'_1,\lambda'_2)F^\Lambda_{\lambda_1\lambda'_1}(\theta_1,\phi_1)G^V_{\lambda_2\lambda'_2}(\theta_2,\phi_2) \quad (11)$$

It is worth noticing that this compact form exhibits clearly the **correlations** between the different angular distributions according to the helicity states of the intermediate resonances :

- $D_{\lambda_1-\lambda_2,\lambda'_1-\lambda'_2}(\theta,\phi) = \sum_{M_i,M'_i} \rho_{M_iM'_i} d^{1/2}_{M_i,\lambda_1-\lambda_2}(\theta) d^{1/2}_{M'_i,\lambda'_1-\lambda'_2}(\theta) \exp i(M_i - M'_i)\phi$

represents a $(4 \times 4)$ hermitean matrix which includes the initial PDM of the $\Lambda_b$.
**Four** hadronic matrix elements are needed to describe the full dynamics of the $\Lambda_b$ decay : $\Lambda_b(1/2,0), \Lambda_b(-1/2,0), \Lambda_b(1/2,1)$ and $\Lambda_b(-1/2,-1)$. Because of parity violation in weak hadronic decays, it is assumed that $\Lambda_b(\lambda_1,\lambda_2) \neq \Lambda_b(-\lambda_1,-\lambda_2)$.

- The two other terms $F^\Lambda_{\lambda_1\lambda'_1}(\theta_1,\phi_1)$ and $G^V_{\lambda_2\lambda'_2}(\theta_2,\phi_2)$ describe the decay dynamics of the intermediate resonances $\Lambda \to P\pi^-$ and $V \to \ell^+\ell^-$ ($h^+h^-$) respectively. They are both hermitean matrices and their explicit forms are given in the appendix : (i) $F^\Lambda$ needs the determination of **two** hadronic matrix elements $\Lambda(1/2,0)$ and $\Lambda(-1/2,0)$ in order to describe the decay $\Lambda \to P\pi^-$. They are related to the proton helicity states and they are unequal because of parity violation. (ii) While for $G^V$, **four** matrix elements must be determined, $V(1/2,1/2)$, $V(-1/2,1/2)$, $V(1/2,-1/2)$ and $V(-1/2,-1/2)$ in the case of leptons and only **one** hadronic element, $V(0,0)$, in the case of final $0^-$ mesons.

## 5 Physical Consequences

The above relations allow us to compute the angular distributions of the three processes taken separately : $\Lambda_b \to \Lambda V$, $\Lambda \to P\pi^-$ and $V \to \ell^+\ell^-$ ($h^+h^-$), by emphasizing the role of the $\Lambda_b$ polarisation.

### 5.1 $\Lambda_b \to \Lambda V$ decay

The angular distributions of this process can be deduced in a straightforward way by integrating the formula displayed in (11) over angles $\theta_1, \phi_1, \theta_2$ and $\phi_2$ and summing over helicities $\lambda_3, \lambda_5$ and $\lambda_6$. Furthermore, the four hadronic matrix elements characterizing $\Lambda_b$ decay can be gathered into two parameters according to the final helicity value :

- $|\Lambda_b(1/2)|^2 = |\Lambda_b(1/2,0)|^2 + |\Lambda_b(-1/2,-1)|^2$

and

- $|\Lambda_b(-1/2)|^2 = |\Lambda_b(-1/2,0)|^2 + |\Lambda_b(1/2,1)|^2$

A first expression arises like

$$\begin{aligned}W(\theta,\phi) &\propto |\Lambda_b(1/2)|^2 \Big(\rho_{++}\cos^2(\theta/2) + \rho_{--}\sin^2(\theta/2) + \Re(\rho_{+-}\exp(-i\phi))\sin\theta\Big) \quad (12)\\ &+ |\Lambda_b(-1/2)|^2 \Big(\rho_{++}\sin^2(\theta/2) + \rho_{--}\cos^2(\theta/2) - \Re(\rho_{+-}\exp(-i\phi))\sin\theta\Big)\end{aligned}$$

Noticing that $\rho_{++} + \rho_{--} = 1$ and $\mathcal{P}_{\Lambda_b} = \rho_{++} - \rho_{--}$, this relation can be modified like the following one :

$$W(\theta,\phi) \propto |\Lambda_b(1/2)|^2 \Big(1 + \mathcal{P}_{\Lambda_b}\cos\theta + 2\Re(\rho_{+-}\exp(-i\phi))\sin\theta\Big) \qquad (13)$$
$$+ |\Lambda_b(-1/2)|^2 \Big(1 - \mathcal{P}_{\Lambda_b}\cos\theta - 2\Re(\rho_{+-}\exp(-i\phi))\sin\theta\Big)$$

Because $|\Lambda_b(1/2)| \neq |\Lambda_b(-1/2)|$, it is clearly seen that parity is violated in the above expression; $W(\theta,\phi) \neq W(\pi-\theta, \pi+\phi)$. This property is put into evidence by introducing the *helicity asymmetry parameter* $\alpha_{As}$ defined by :

$$\alpha_{AS} = \frac{|\Lambda_b(1/2)|^2 - |\Lambda_b(-1/2)|^2}{|\Lambda_b(1/2)|^2 + |\Lambda_b(-1/2)|^2} \qquad (14)$$

The final angular distribution will be expressed like :

$$W(\theta,\phi) \propto 1 + \alpha_{AS}\mathcal{P}_{\Lambda_b}\cos\theta + 2\alpha_{AS}\Re(\rho_{+-}\exp(-i\phi))\sin\theta \qquad (15)$$

Then, averaging over the azimuthal angle $\phi$, a *standard* formula is obtained for the polar angle distribution :

$$\frac{dN}{\cos\theta} \propto 1 + \alpha_{AS}\mathcal{P}_{\Lambda_b}\cos\theta \qquad (16)$$

which is very similar to the proton angular distribution in the $\Lambda \to P\pi^-$ decay [8].

• It could be noticed that the angular dissymetries which arise in the $\Lambda_b$ decays are possible *only if the initial polarization* $\mathcal{P}_{\Lambda_b} \neq 0$.

## 5.2 $\Lambda \to P\pi^-$ decay and its Polarization Density Matrix

This process represents a "standard" weak hadronic two-body decay. Although its decay probability is very known in the literature, our aim is to shed a new light on the PDM of the $\Lambda \to P\pi^-$ decay, process related to $\mathcal{P}_{\Lambda_b}$.

Departing from relation (11), integrating over the angles $\theta, \phi, \theta_2, \phi_2$ and summing over vector-meson helicity states, a general formula for proton angular distributions in the $\Lambda$ frame can be recovered :

$$W_1(\theta_1, \phi_1) \propto K_1\Big(|\Lambda(+)|^2\cos^2(\theta_1/2) + |\Lambda(-)|^2\sin^2(\theta_1/2)\Big) \qquad (17)$$
$$+ K_2\Big(|\Lambda(+)|^2\sin^2(\theta_1/2) + |\Lambda(-)|^2\cos^2(\theta_1/2)\Big)$$
$$- \frac{\pi}{8}\mathcal{P}_{\Lambda_b}K_3\exp(i\phi)\big(|\Lambda(+)|^2 - |\Lambda(-)|^2\big)\sin\theta_1$$
$$- \frac{\pi}{8}\mathcal{P}_{\Lambda_b}K_3^*\exp(-i\phi)\big(|\Lambda(+)|^2 - |\Lambda(-)|^2\big)\sin\theta_1$$

where $|\Lambda(+)| = |\Lambda(1/2)|$, $|\Lambda(-)| = |\Lambda(-1/2)|$ and $K_1, K_2, K_3$ are analytical expressions obtained from the matrix elements $D_{M_f M'_f}$ and $G_{ij}^V$ defined above.

By arranging again the different terms and using the angle $\theta_1$ instead of $\theta_1/2$ , a general formula is obtained for the proton angular distributions :

$$\begin{aligned}W_1(\theta_1,\phi_1) \propto\ & \frac{(K_1-K_2)}{2}\bigl(|\Lambda(+)|^2-|\Lambda(-)|^2\bigr)\cos\theta_1 \\ & - \frac{\pi}{4}\mathcal{P}_{\Lambda_b}\bigl(|\Lambda(+)|^2-|\Lambda(-)|^2\bigr)\Re(K_3\exp(i\phi_1))\sin\theta_1 \\ & + \frac{(K_1+K_2)}{2}\bigl(|\Lambda(+)|^2+|\Lambda(-)|^2\bigr)\end{aligned} \quad (18)$$

- This relation shows clearly that the initial $\Lambda_b$ polarization arises again in the *interference term* relating the polar angle $\theta_1$ and the azimuthal angle $\phi_1$.
- The PDM elements of the $\Lambda$ can be computed from the above relation. By performing standard calculations of a polarized $\Lambda$ decaying into $P\pi^-$ , we deduce (to a normalization factor) :

$$\rho^\Lambda_{++} = K_1 = \int G^V_{00}\,|\Lambda_b(1/2,0)|^2 + \int G^V_{11}\,|\Lambda_b(1/2,1)|^2 \quad (19)$$

$$\rho^\Lambda_{--} = K_2 = \int G^V_{00}\,|\Lambda_b(-1/2,0)|^2 + \int G^V_{-1-1}\,|\Lambda_b(-1/2,-1)|^2 \quad (20)$$

$$\rho^\Lambda_{+-} = -\frac{\pi}{4}\mathcal{P}_{\Lambda_b}\int G^V_{00}\,\Lambda_b(-1/2,0)\Lambda_b^*(1/2,0) \quad (21)$$

Here $\int G^V_{ii}$ $(i=-1,0,1)$ designates the integral of the functions $G^V_{ii}$ over angles $\theta_2$ and $\phi_2$.

## 5.3  $V \to \ell^+\ell^-\ (h^+h^-)$ decays

Vector meson V decaying into a lepton pair or a hadronic one is described by the $(3\times 3)$ hermitean matrix $G^V$. The analytic expressions of the $G^V$ matrix elements change according to the spin of the final particles (leptons or pseudoscalar mesons). The angular distributions $W_2(\theta_2,\phi_2)$ in the V rest-frame are obtained by integrating relation (11) over the angles $\theta,\phi,\theta_1,\phi_1$ and summing over the two $\Lambda$ helicity states :

$$\begin{aligned}W_2(\theta_2,\phi_2) \propto\ & \\ & \bigl(|\Lambda_b(1/2,0)|^2+|\Lambda_b(-1/2,0)|^2\bigr)G^V_{00} + |\Lambda_b(1/2,1)|^2 G^V_{11} + |\Lambda_b(-1/2,-1)|^2 G^V_{-1-1} \\ & - \frac{\pi}{4}\mathcal{P}_{\Lambda_b}\Bigl(\Lambda_b(1/2,0)\Lambda_b^*(1/2,1)G^V_{01} + \Lambda_b(1/2,1)\Lambda_b^*(1/2,0)G^V_{10} \\ & + \Lambda_b(-1/2,0)\Lambda_b^*(-1/2,-1)G^V_{0-1} + \Lambda_b(-1/2,-1)\Lambda_b^*(-1/2,0)G^V_{-10}\Bigr)\end{aligned} \quad (22)$$

Again, the initial $\Lambda_b$ polarization appears explicitely in the interference terms (non-diagonal $G^V_{ij}$ matrix elements) relating the angles $\theta_2$ and $\phi_2$. According to the analytic forms of $G^V_{ij}$ , the following relation for $W_2(\theta_2,\phi_2)$ is obtained :

$$W_2(\theta_2,\phi_2) \propto A\cos^2\theta_2 + B\sin^2\theta_2 + C - \frac{\pi}{4}\mathcal{P}_{\Lambda_b}\Re\Bigl(D\exp(i\phi_2)\sin(2\theta_2)\Bigr) \quad (23)$$

Averaging over the azimuthal angle $\phi_2$, the $\theta_2$ polar angle distribution can be deduced :

Case of leptonic decays

$$\frac{dN}{d\cos\theta_2} \propto (1 - 3\rho_{00}^V)\cos^2\theta_2 \; + \; (1 + \rho_{00}^V) \qquad (24)$$

Case of hadronic decays

$$\frac{dN}{d\cos\theta_2} \propto (3\rho_{00}^V - 1)\cos^2\theta_2 \; + \; (1 - \rho_{00}^V) \qquad (25)$$

where $\rho_{00}^V$ is PDM matrix element indicating the rate of longitudinal polarization of the vector meson V.

N.B : Since the vector meson V decays essentially via Strong or Electromagnetic Interactions, *parity* is always supposed conserved in the computation of the decay amplitudes; which is clearly seen from the angular distributions shown above.

## 5.4 Three Body Decays

These calculations can be extended to the three body decays like $\Lambda_b \to \Lambda \ell^+ \ell^-$ or $\Lambda h^+ h^-$ where the lepton or meson pair does not originate from a resonance with a given mass, but respectively from a virtual photon (Electroweak penguin diagram) or a virtual gluon (QCD penguin diagram). In this case, an additionnal parameter arises, which is the *invariant mass* of the $(\ell^+\ell^-)$ or $(h^+h^-)$ system denoted by $\sqrt{Q_{lep}^2}$ or $\sqrt{Q_{had}^2}$. In our simulations and in accordance with the spirit of this paper, this mass is generated uniformly in the interval $[M_{min}, M_{max}]$ with $M_{min} = 2m_{l(h)}$ and $M_{max} = m(\Lambda_b) - m(\Lambda)$ ; the computations of the corresponding angular distributions are still *identical to the preceding case* where the vector meson mass is fixed.

In the next paper [5] and by means of the OPE techniques, the $Q_{lep(had)}^2$ can be derived and we expect to obtain all the spectra of the main kinematic variables which characterize the three body decays.

## 6 Preliminary Conclusions

• Complete calculations of the angular distributions of the process $\Lambda_b \to \Lambda V$ , $\Lambda \to P\pi^-$ and $V \to \ell^+\ell^-$ have been performed by using the helicity formalism and stressing on the correlations which arise among the final decay products.

• In all these calculations, particular role of the $\Lambda_b$ polarization has been put into evidence. $\mathcal{P}_{\Lambda_b}$ appears *explicitely* in the polar angle distribution of the $\Lambda$ hyperon in the $\Lambda_b$ restframe and in the azimuthal angle distributions of both proton and $\ell^-$ respectively in the $\Lambda$ and V frames. Averaging over the angles $\phi_1$ and $\phi_2$ according to the quantization axis annihilates the effects of $\mathcal{P}_{\Lambda_b}$.

• Polarization Density Matrices of $\Lambda$ hyperon and vector meson V can be inferred from the general formula shown in relation (11). Explicit calculations of $\rho^\Lambda$ and $\rho^V$ matrix elements require the computation of several hadronic matrix elements via the OPE techniques and the knowledge of several form factors.

These exhaustive calculations and the full kinematics simulations of the decay processes are the subject of a forthcoming publication.

## Acknowledgements

Z.J.A. has benefited from the advices and the skillness of Dr C.Rimbault Both the two authors are very indebted to their colleagues of the LHCB Clermont-Ferrand team and especially to Dr P.Perret for his continuous interest and his encouragements in developing this new and very promising research subject of Time-Reversal symmetry.

# Appendix

## A  Analytical expression of $\Lambda \to P\pi^-$ decay matrix

The departure formula is the following one :

$$F^{\Lambda}_{\lambda_1 \lambda'_1}(\theta_1, \phi_1) = \exp i(\lambda_1 - \lambda'_1)\phi_1 \Big(|\Lambda(+)|^2 d^{1/2}_{\lambda_1 1/2}(\theta_1) d^{1/2}_{\lambda'_1 1/2}(\theta_1) + |\Lambda(-)|^2 d^{1/2}_{\lambda_1 -1/2}(\theta_1) d^{1/2}_{\lambda'_1 -1/2}(\theta_1)\Big)$$

where $|\Lambda(+)|$ and $|\Lambda(-)|$ are defined in the text. In the following, the angular arguments of $F^{\Lambda}_{\lambda_1 \lambda'_1}$ and those of the Wigner matrix elements $d^j_{mm'}$ will be omitted in order to simplify the notations.

$$F^{\Lambda}_{1/2,1/2} = |\Lambda(+)|^2 \cos^2(\theta_1/2) + |\Lambda(-)|^2 \sin^2(\theta_1/2) \qquad (26)$$

$$F^{\Lambda}_{1/2,-1/2} = \frac{1}{2}\Big(|\Lambda(+)|^2 - |\Lambda(-)|^2\Big) \sin\theta_1 \exp(i\phi_1) \qquad (27)$$

$$F^{\Lambda}_{-1/2,1/2} = \frac{1}{2}\Big(|\Lambda(+)|^2 - |\Lambda(-)|^2\Big) \sin\theta_1 \exp(-i\phi_1) \qquad (28)$$

$$F^{\Lambda}_{-1/2,-1/2} = |\Lambda(+)|^2 \sin^2(\theta_1/2) + |\Lambda(-)|^2 \cos^2(\theta_1/2) \qquad (29)$$

It can be noticed that, by integrating over the angular ranges of $\theta_1$ and $\phi_1$, the non-diagonal terms disappear and only the diagonal ones remain in the final calculations.

## B  Analytical expression of $V \to \ell^+\ell^-$ $(h^+h^-)$ decay matrix

The $(3 \times 3)$ hermitean matrix $G^V$ describing the above process has the following matrix elements :

$$G^V_{\lambda_2 \lambda'_2}(\theta_2, \phi_2) = \sum_{\lambda_5, \lambda_6} |V(\lambda_5, \lambda_6)|^2 d^1_{\lambda_2 m_2}(\theta_2) d^1_{\lambda'_2 m_2}(\theta_2) \exp i(\lambda_2 - \lambda'_2)\phi_2$$

where $m_2 = \lambda_5 - \lambda_6$. The decay matrix of the channel $\rho^0 \to \pi^+\pi^-$ will be given explicitely. In this case, $m_2 = m'_2 = 0$ and only one form factor related to the hadronic matrix element $\rho(0,0)$ is needed to complete the decay dynamics. It will be omitted in the following expressions :

$$G^\rho_{11} = \frac{\sin^2\theta}{2} \ , \ G^\rho_{10} = \frac{-\sin 2\theta \exp(i\phi)}{2\sqrt{2}} \ , \ G^\rho_{1-1} = \frac{-\sin^2\theta \exp(2i\phi)}{2} \tag{30}$$

$$G^\rho_{01} = \frac{-\sin 2\theta \exp(-i\phi)}{2\sqrt{2}} \ , \ G^\rho_{00} = \cos^2\theta \ , \ G^\rho_{0-1} = \frac{\sin 2\theta \exp(i\phi)}{2\sqrt{2}} \tag{31}$$

$$G^\rho_{-11} = \frac{-\sin^2\theta \exp(-2i\phi)}{2} \ , \ G^\rho_{-10} = \frac{\sin 2\theta \exp(-i\phi)}{2\sqrt{2}} \ , \ G^\rho_{-1-1} = \frac{\sin^2\theta}{2} \tag{32}$$

In the case of $V \to \ell^+\ell^-$ with *parity conservation*, the analytical expressions of $G^V_{ij}$ are considerably simplified because $|V(\lambda_5,\lambda_6)| = |V(-\lambda_5,-\lambda_6)|$. Departing from these considerations, the following expressions are obtained :

$$G^V_{00} = 2|V(++)|^2 \cos^2\theta + |V(+-)|^2 \sin^2\theta \tag{33}$$

$$G^V_{11} = G^V_{-1-1} = |V(++)|^2 \sin^2\theta + \frac{1}{2}|V(+-)|^2(1+\cos^2\theta) \tag{34}$$

$$G^V_{01} = (G^V_{10})^* = \left(|V(+-)|^2 - 2|V(++)|^2\right) \frac{\sin 2\theta \exp(-i\phi)}{2\sqrt{2}} \tag{35}$$

$$G^V_{0-1} = -G^V_{10} = \left(2|V(++)|^2 - |V(+-)|^2\right) \frac{\sin 2\theta \exp(i\phi)}{2\sqrt{2}} \tag{36}$$

$$G^V_{-10} = (G^V_{0-1})^* = \left(2|V(++)|^2 - |V(+-)|^2\right) \frac{\sin 2\theta \exp(-i\phi)}{2\sqrt{2}} \tag{37}$$

Like the previous case, integrating over the angular ranges of $\theta_2$ and $\phi_2$ annihilates the non-diagonal terms which are due to interferences among different helicity states.

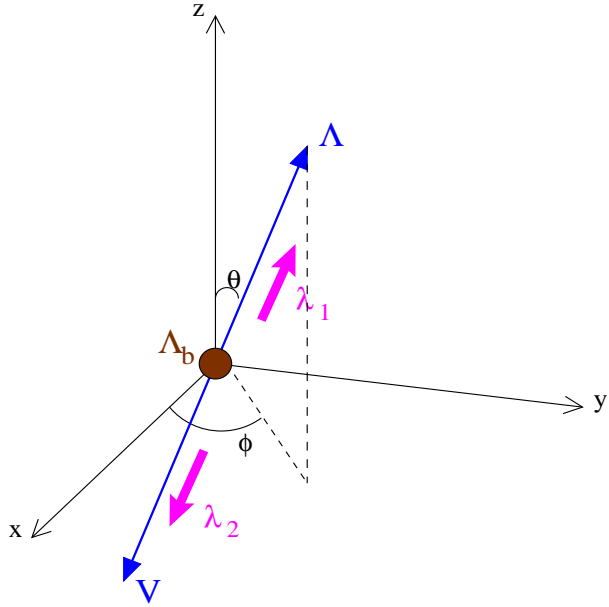

Figure 1: $\Lambda_b$ *decay in its transversity frame*

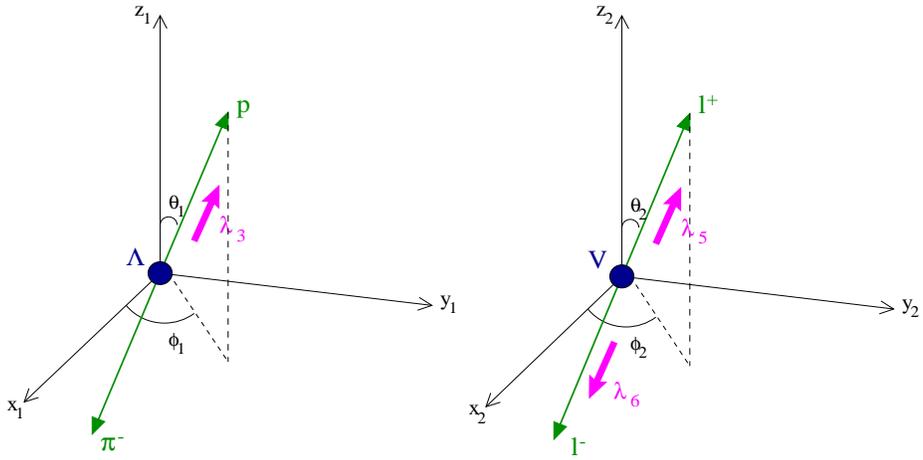

Figure 2: *Helicity frames respectively for $\Lambda$ and Vector Meson V decays*